# Optical Post Processing for High Speed Quantum Random Number Generators

Abdulrahman Dandasi,[1] Helin Ozel,[1,2] Orkun Hasekioglu,[2] and Kadir Durak[1,2]

[1]Quantum Optics Laboratory, Graduate School of Engineering and Science, Electrical and Electronics Department. Ozyegin University, Istanbul 34794, Turkey

[2]Informatics and Information Security Research Center BILGEM, Scientific and Technological Research Council of Turkey TUBITAK, Istanbul 41400, Turkey.

The speed of quantum random number generators is a major concern for practical quantum applications. However, the bit extraction process limits the final bit rate due to lack of comparably fast electronics. Here we introduce optical scattering as a method to perform optical bit extraction. Scattering is a probabilistic phenomenon and it increases the chaotic behaviour of coherent sources. As a result, it broadens the distribution of photon statistics and makes it super-Poissonian. We show that the raw signal of the sources with super-Poissonian distribution have better randomness compared to Poissonian, indicated by their autocorrelation characteristics. Therefore, the optical bit extraction process allows faster sampling of raw signal without compromising the randomness quality. The use of scattering mechanisms as an entropy source eases the miniaturization of quantum random number generators, it also makes them compatible and adaptable to existing technologies.

*Introduction*.—The importance of random number generators (RNGs) is increasing day by day, alongside the advancement of technology and science. Nowadays, there is a wide range of applications for random number generators such as computer games, simulations, lotteries, statistics, financial transactions and communication security [1-5]. For most of these applications, random numbers are required at a high rate [6]. Pseudorandom-number generators, that are based on complex mathematical algorithms, can provide high bit rate. However, they fail to be truly random as there exist algorithms that can examine the pattern of the PRNG and predict the next coming bits.

Quantum Random Number Generators (QRNGs), on the other hand, are proved to be the most secure ones because their entropy source is completely random and unpredictable. There are reports on QRNGs based on Raman scattering [7], vacuum fluctuations [8-11], optical vacuum states [12-17], photon arrival time [18-22], single photon detection [23, 24] spontaneous emission noise [25-31], photon bunching [32], uncertainty principle [33], photon number distribution [34, 35], photon polarization state [36, 37], mobile phone camera [38].

Ideal cryptography is based on one-time-pad encryption requires key generation and distribution at the same rate as the data transmission rate [39]. Although it is possible to generate keys at very high rates using optical QRNG, typically the optical signals are converted to electrical signals through detectors for post processing. The post processing rate is around two orders of magnitude slower than the optical signal generation rate [40]. Therefore, it is crucial to have the capability to perform post processing (bit extraction) algorithms optically to achieve the targeted final bit rate.

The random nature of quantum phenomena makes it an appealing candidate for RNGs. Vacuum fluctuations, quantum tunnelling and interference are some of the events that are used in the literature for quantum random number generation. The shot noise based QRNGs are shown to be fast and reliable thanks to their unpredictable entropy source; vacuum fluctuations. Tunnelling or scattering effects also allow fast and reliable creation of quantum random numbers.

The photon statistics of a coherent light source is Poissonian. The scattering of the coherent light by the scattering centers lead the distribution to have a greater variance. Consequently, the distribution turns into a super-Poissonian where the fluctuations in the measured optical intensity have a broader distribution even with the same mean value. The autocorrelation function of the raw signal for a coherent light and scattered light show different lag values for minimum (ideally zero) values. We show that for super-Poissonian the minimum of autocorrelation function occurs at a smaller lag value. This allows a faster sampling rate without compromising the randomness quality.

*Experiment*.—We classify the light sources based on the variance of their photon distributions. They are categorized into three main types; Poissonian, sub-Poissonian, and super-Poissonian [41]. The super-Poissonian is either chaotic, incoherent, or thermal light. Where the Poissonian distribution is caused by a perfectly coherent light, and the sub-Poissonian distribution is caused by a non-classical light source. Despite the fact that they all share the same mean, the distributions of super-Poissonian and sub-Poissonian look broader and narrower than the Poissonian, respectively.

In the literature, many of the applications have been implemented on the balanced homodyne detection to monitor the quadrature measurements of the quantum vacuum fluctuations. The main reason behind its preferrasion is to remove the effect of the classical noise. However in our case (see Fig.1), classical noise wanted to be observed in order to show the dominance of the quantum noise that is why homodyne detection is not preferred.

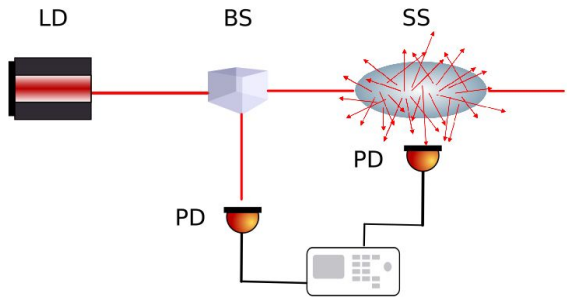

Fig.1: Schematic diagram of scattering based RNG. The laser diode (LD), Beam Sampler (BS), Silver oval surface in 5mm diameter is used as scattering source (SS). Photo detectors (PD) are used to compare the scattered rays.

For this experiment NKT Koheras Basic E15 laser with the central wavelength of 1550 nm and linewidth of 100 Hz is used for different scattering mechanisms like; unpolished surfaces, such as aluminum, steel, gold and silver.

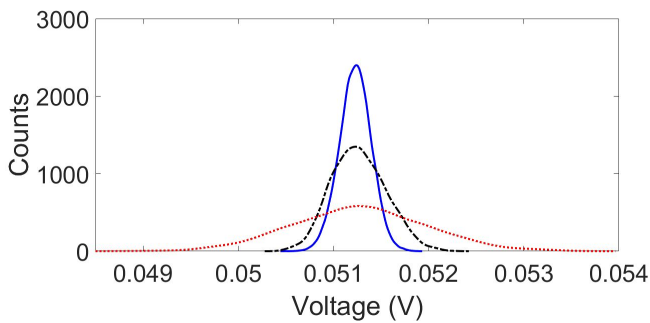

Fig.2: Density function of the laser distribution shows a Poisson distribution as expected based on being a coherent light having a FWHM of 0.40 mV. The scattered laser from a half-polished silver surface shows a Super-Poissonian distribution with a FWHM of 0.67 mV, while the scattering from an unpolished Aluminum surface has a FWHM of 1.67 mV.

While different scattering sources have different scattering parameters, all result in a broadened distribution of the photon statistics. Fig.2 shows the measured distributions of the light from a coherent laser, scattered light from half-polished silver and unpolished aluminum.

The light from the laser exhibits Poissonian distribution. The random nature of scatterings from a semi-polished silver mirror enhances the chaotic (thermal) characteristics of the light and broadens the distribution. Unpolished aluminum has higher intensity of scattering centers and the roughness of the surface is proportional to the scattering. Therefore the distribution is further broadened. This shows that the chaotic behaviour becomes more dominant with the increased number of scattering centers.

*Min-entropy estimation.*—One crucial aspect of proving the randomness of a stream of bits is how much entropy it can produce per output bit. The perfect entropy in a system exists only if the number of 0's and 1's is exactly equal in a bit stream. Where each bit can be used as an independent randomness where it resembles a fair coin toss. Min-entropy is the most moderate bound of a functional entropy for a randomness source, where it is not feasible to well quantify the randomness by other entropies such as Shannon Entropy [42, 43]. In order to ensure that the randomness we are extracting is truly quantum, and not just electronic noise, we need to understand each noise source separately. This is necessary to insure the validity of the extracted quantum noise and predict its entropy.

Min-entropy is used to quantify the amount of randomness in the system. Where by using min-entropy we will be able to decide on how many bits to be received from each set of samples. In the process of calculating min-entropy, we will make a few assumptions; that the signal detected by the detector is the sum of two different noises; quantum noise (shot noise) and electronic noise (technical noise) given by $X_t = X_q + X_e$ and that they are both independent of each other [38]. Also that the photon statistic of the quantum signal will follow a Poisson distribution, and that the classical noise will show a Gaussian distribution [41]. The last assumption is the worst case-scenario is that an adversary (Eve) has full knowledge of the classical noise. Both the quantum signal and the classical noise will be digitized as voltage values by an oscilloscope. The formula to calculate min-entropy is:

$$H_\infty(X) = -\log(P[X = x]) \quad (1)$$

where $H_\infty(X)$ is the min-entropy of a probability distribution X on {0,1}, and n is the count of the used bits in digitizing the signal into voltages. And $P[X = x]$ is the probability that the measured voltage V will take place in bin X. Out of 16 bits, we get a min-entropy of 13.45 bits per sample. Which is the maximum entropy an eavesdropper can guess from a random sample of our data. Similar to the total noise. The total variance $\sigma^2$ equals the sum of the thermal variance $\sigma_e^2$ and the quantum variance $\sigma_q^2$ [44]. According to [41], you can calculate the shot noise level by using the following equation:

$$P_{quantum} = 2eR_L i\Delta f \qquad (2)$$

Where e corresponds to the modulus of the charge of the electron, $R_L$ is the load resistance, i is the photocurrent and $\Delta f$ as the bandwidth frequency. For a 1 mW power set of samples, the theoretical shot noise is calculated as -69.2 dBm.

There are some limitations in the literature such as the frequency bandwidth and the dead time of the detector [8]. Using 5 GHz frequency bandwidth detectors alleviates the limitation. The lower bound is determined by the min-entropy as it is explained above. In our paper most challenges and limitations originate from classical noise and post processing [40]. In Fig.3, shot noise is dominant to classical noise so what is left is the post processing part. We created an alternative to that by using scattering as a way to do optical bit extraction.

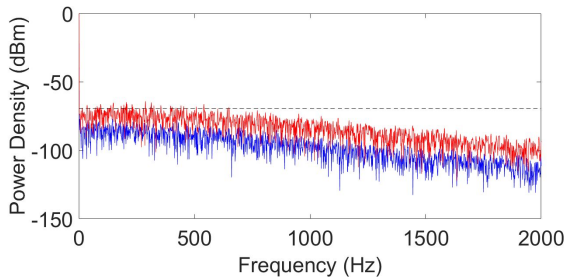

Fig. 3: Power spectral density of the measured signal (red) vs electronic noise (blue). Large fluctuations are observed, wherefore the high acquisition rate of the oscilloscope. The black dotted line shows the theoretically calculated shot noise level.

Power spectral density of the signal is measured to guarantee that the quantum fluctuations dominate the electrical noise.

Randomness *extraction.*—Randomness extraction is a method to elicit a bunch of bits in the shape of a distribution that is almost uniform coming from biased and correlated bits. Originally, the main purpose of these extraction methods was to randomize the data of a weak randomness source more. This purpose is still attracting to be used for the same reason, but recently this method had gained notable importance especially in the field of cryptography and computer science. It is also important because when we smooth the entropy source, Shannon entropy function is maximized. There are different methods in the literature to extract randomness, some use a non-fuzzy randomness extractor as in [45]. Another popular method is entropy hashing method [7, 46]. Some others use Common Chaining Modes [47], another one is the Permutation Based Extractors [48].

In this paper, the implemented randomness extraction method is the Linear-feedback shift register (LFSR). It is used to find the shortest linear feedback register which can generate a given finite sequence of digits [49]. LFSRs are also operating as a pseudorandom number generator [50].

*Results.*—There exists a few methods to determine the randomness quality of the QRNG. One of them is the autocorrelation function of bit stream as it shows the unpredictability of the next bit from the bit before it.

In order to suppress the effects of electronic noise, we sample our data at a sampling rate of 20 GSamples/s. When the sampling rate is higher, the correlation between the next sampling point is also higher. Hence, to determine the sampling rate the autocorrelation coefficient of the output signal needs to be calculated at assorted sampling frequencies, $f_s = 1/T_s$.

The sampling rate of the oscilloscope is 40 GSps, which is much faster than the bandwidth of our detector (5 GHz). Therefore, the signal requires undersampling. The ideal undersampling value can be extracted from the autocorrelation function. The autocorrelations of the samples, shape the primary approximation for the bit sequence quality. The minimum autocorrelation in the signal provides better randomness. If the minimum autocorrelation occurs at small lag values the sampling rate can be faster and therefore a higher final bit rate can be achieved. The autocorrelation function of the scattering comes around zero faster than the direct laser signal (see Fig.4), this lag point is used for undersampling.

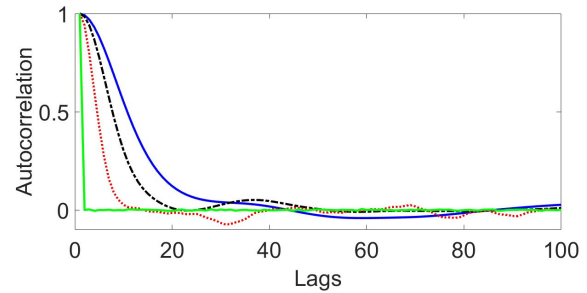

Figure 4: Autocorrelation functions of the analog signal from the laser directly (blue), after scattering half polished silver surface (black), after scattering of unpolished aluminum (red), after and the laser after digital bit extraction process (green).

For the scattered source the minimum autocorrelation value is much smaller than the analog signal from the laser fluctuations. However, the digital bit extraction (we used LFSR in this work) requires smaller lag value for minimum autocorrelation compared to the optical scattering case. The bit extraction process is used for signal probability equalization and decreased self-correlations. Combining this graph with density functions of these three signals (see Fig. 2) it can be concluded that the optical scattering works as a bit extraction process. However, there is still room for improvement on the ideal optical bit extraction process as the digital bit extraction process results in better signal equalization and allows faster sampling rate. The lower value of the autocorrelation function coefficient in scattered data compared to direct laser implies that the higher quality randomness can be obtained with scattering.

Where the direct laser becomes negative at lag 44, half polished silver at 22, unpolished aluminium at lag 15, and after only 2 lag points for the bit extracted laser.

*Conclusion.*—This paper introduces a technique for a QRNG based on light scattering. The method is used to perform a fast post processing with optical bit extraction.

Many constraints are encountered in the physical production of random numbers; the bandwidth and the deadtime of the detector, classical noise, digitizing of the analog data and the post processing are the big obstacles ahead that slow down QRNG. Our method of using scattering source, proposes a solution for this problem by giving the physical setup the ability to produce data in a flatter and more random.

The photon statistics of thermal and coherent light sources and their scatterings are compared. The sources with super-Poissonian distribution are shown to have better randomness than the Poisson distribution. Moreover, the autocorrelation function of the scattering signal gets around zero in advance compared to the coherent laser signal. The optimal sampling rate is obtained when the undersampling is set as two times of the lag point for the minimum autocorrelation of the scattered signal. The raw signal is digitized and processed optically in order to make an even distribution of bits.

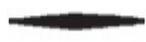

[1] S. L. Lohr, Sampling: Design and Analysis (Duxbury Press, Florence, 1999).
[2] Sumanjeet, 2009 First Asian Himalayas International Conference on Internet (2009).
[3] N. Metropolis and S. Ulam, J. Am. Stat. Assoc. 44, 335 (1949).
[4] R. L. Rivest, A. Shamir, and L. Adleman, (1978).
[5] W. Diffie and M. Hellman, IEEE Transactions on Information Theory **22**, 644 (1976).
[6] N. Gisin, G. Ribordy, W. Tittel, and H. Zbinden, Reviews of Modern Physics **74**, 145 (2002).
[7] P. J. Bustard, D. G. England, J. Nunn, D. Moffatt, M. Spanner, R. Lausten, and B. J. Sussman, Optics Express **21**, 29350 (2013).
[8] Y. Shi, B. Chng, and C. Kurtsiefer, Applied Physics Letters **109**, 041101 (2016).
[9] T. Symul, S. M. Assad, and P. K. Lam, Applied Physics Letters **98**, 231103 (2011).
[10] Q. Zhou, R. Valivarthi, C. John, and W. Tittel, Quantum Engineering **1**, (2019).
[11] Z. Zheng, Y. Zhang, W. Huang, S. Yu, and H. Guo, Review of Scientific Instruments **90**, 043105 (2019).
[12] J. Y. Haw, S. M. Assad, A. M. Lance, N. H. Y. Ng, V. Sharma, P. K. Lam, and T. Symul, Physical Review Applied **3**, (2015).
[13] F. Raffaelli, G. Ferranti, D. H. Mahler, P. Sibson, J. E. Kennard, A. Santamato, G. Sinclair, D. Bonneau, M. G. Thompson, and J. C. F. Matthews, Quantum Science and Technology **3**, 025003 (2018).
[14] C. Gabriel, C. Wittmann, D. Sych, R. Dong, W. Mauerer, U. L. Andersen, C. Marquardt, and G. Leuchs, Nature Photonics **4**, 711 (2010).
[15] B. Xu, Z. Chen, Z. Li, J. Yang, Q. Su, W. Huang, Y. Zhang, and H. Guo, Quantum Science and Technology **4**, 025013 (2019).
[16] Y. Shen, L. Tian, and H. Zou, Physical Review A **81**, (2010).
[17] C. Gabriel, C. Wittmann, D. Sych, R. Dong, W. Mauerer, U. L. Andersen, C. Marquardt, and G. Leuchs, Nature Photonics **4**, 711 (2010).
[18] M. Wahl, M. Leifgen, M. Berlin, T. Röhlicke, H.-J. Rahn, and O. Benson, Applied Physics Letters **98**, 171105 (2011).
[19] H.-Q. Ma, Y. Xie, and L.-A. Wu, Applied Optics **44**, 7760 (2005).
[20] M. A. Wayne, E. R. Jeffrey, G. M. Akselrod, and P. G. Kwiat, Journal of Modern Optics **56**, 516 (2009).
[21] M. A. Wayne and P. G. Kwiat, Optics Express **18**, 9351 (2010).
[22] J.-M. Wang, T.-Y. Xie, H.-F. Zhang, D.-X. Yang, C. Xie, and J. Wang, IEEE Photonics Journal **7**, 1 (2015).
[23] J. F. Dynes, Z. L. Yuan, A. W. Sharpe, and A. J. Shields, Applied Physics Letters **93**, 031109 (2008).
[24] Y.-Q. Nie, H.-F. Zhang, Z. Zhang, J. Wang, X. Ma, J. Zhang, and J.-W. Pan, Applied Physics Letters **104**, 051110 (2014).

[25] Y. Liu, M.-Y. Zhu, B. Luo, J.-W. Zhang, and H. Guo, Laser Physics Letters **10**, 045001 (2013).
[26] A. Martin, B. Sanguinetti, C. C. W. Lim, R. Houlmann, and H. Zbinden, Journal of Lightwave Technology **33**, 2855 (2015).
[27] X. Li, A. B. Cohen, T. E. Murphy, and R. Roy, Optics Letters **36**, 1020 (2011).
[28] Stipčević M. and B. M. Rogina, Review of Scientific Instruments **78**, 045104 (2007).
[29] C. R. S. Williams, J. C. Salevan, X. Li, R. Roy, and T. E. Murphy, Optics Express **18**, 23584 (2010).
[30] B. Qi, Y.-M. Chi, H.-K. Lo, and L. Qian, Optics Letters **35**, 312 (2010).
[31] W. Wei, G. Xie, A. Dang, and H. Guo, IEEE Photonics Technology Letters **24**, 437 (2012).
[32] Y. Guo, C. Peng, Y. Ji, P. Li, Y. Guo, and X. Guo, Optics Express **26**, 5991 (2018).
[33] G. Vallone, D. G. Marangon, M. Tomasin, and P. Villoresi, Physical Review A **90**, (2014).
[34] H. P. Yuen and V. W. S. Chan, Optics Letters **8**, 177 (1983).
[35] E. Jakeman, C. Oliver, and E. Pike, Advances in Physics **24**, 349 (1975).
[36] M. Fiorentino, C. Santori, S. M. Spillane, R. G. Beausoleil, and W. J. Munro, Physical Review A **75**, (2007).
[37] T. Jennewein, U. Achleitner, G. Weihs, H. Weinfurter, and A. Zeilinger, Review of Scientific Instruments **71**, 1675 (2000).
[38] B. Sanguinetti, A. Martin, H. Zbinden, and N. Gisin, Physical Review X **4**, (2014).
[39] F.-G. Deng and G. L. Long, Physical Review A **69**, (2004).
[40] M. Herrero-Collantes and J. C. Garcia-Escartin, Reviews of Modern Physics **89**, (2017).
[41] M. Fox, *Quantum Optics: an Introduction* (2006).
[42] B. Chor and O. Goldreich, 26th Annual Symposium on Foundations of Computer Science (Sfcs 1985).
[43] D. Zuckerman, Proceedings [1990] 31st Annual Symposium on Foundations of Computer Science.
[44] F. Xu, B. Qi, X. Ma, H. Xu, H. Zheng, and H.-K. Lo, Optics Express **20**, 12366 (2012).
[45] N. Nisan and D. Zuckerman, Journal of Computer and System Sciences **52**, 43 (1996).
[46] R. Impagliazzo and M. Luby, 30th Annual Symposium on Foundations of Computer Science (1989).
[47] Y. Dodis, R. Gennaro, J. Håstad, H. Krawczyk, and T. Rabin, Advances in Cryptology – CRYPTO 2004 Lecture Notes in Computer Science 494 (2004).
[48] J. J. M. Chan, P. Thulasiraman, G. Thomas, and R. Thulasiram, Journal of Computer and Communications **04**, 73 (2016).
[49] J. Massey, IEEE Transactions on Information Theory **15**, 122 (1969).
[50] M. Saito and M. Matsumoto, Monte Carlo and Quasi-Monte Carlo Methods 2006 607 (2008).